\documentclass[journal]{IEEEtran}
\usepackage{subscript}
\usepackage{fixltx2e}

\usepackage{graphicx,booktabs} 
\usepackage{url,hyperref,lineno,microtype,subcaption}
\usepackage{float}
\usepackage{verbatim} 
\usepackage{multirow}
\usepackage{booktabs}
\usepackage{textcomp}
\usepackage{siunitx}
\usepackage{amsmath}
 \usepackage{enumitem}
 \usepackage{makecell}
 \usepackage{siunitx}
 \usepackage{amsfonts}
\usepackage{gensymb}
\usepackage{enumitem}
\usepackage{cleveref}
\usepackage[table,xcdraw]{xcolor}

\usepackage{tikz}
\usepackage{textcomp}
\usepackage{hyperref}
\usepackage{lipsum}

\newcommand\copyrighttext{%
  \footnotesize \textcopyright 2023 IEEE. Personal use of this material is permitted.
  Permission from IEEE must be obtained for all other uses, in any current or future 
  media, including reprinting/republishing this material for advertising or promotional 
  purposes, creating new collective works, for resale or redistribution to servers or 
  lists, or reuse of any copyrighted component of this work in other works. 
 }
\newcommand\copyrightnotice{%
\begin{tikzpicture}[remember picture,overlay]
\node[anchor=south,yshift=10pt] at (current page.south) {\fbox{\parbox{\dimexpr\textwidth-\fboxsep-\fboxrule\relax}{\copyrighttext}}};
\end{tikzpicture}%
}

\usepackage{cite}

\ifCLASSINFOpdf
\else
\fi
\IEEEoverridecommandlockouts

\begin{document}


%
\title{
Evaluation of short range depth sonifications for visual-to-auditory sensory substitution
}

%
%
%

\author{Louis~Commère,
        and~Jean~Rouat,~\IEEEmembership{Senior,~IEEE}
\thanks{
}
\thanks{
}
}

%
%

\markboth{Journal of \LaTeX\ Class Files,~Vol.~14, No.~8, August~2015}%
{Shell \MakeLowercase{\textit{et al.}}: Bare Demo of IEEEtran.cls for IEEE Journals}
%



\maketitle

\copyrightnotice

\begin{abstract}
Visual to auditory sensory substitution devices convert visual information into sound and can provide valuable assistance for blind people.
Recent iterations of these devices rely on depth sensors. 
Rules for converting depth into sound (i.e. the sonifications) are often designed arbitrarily, with no strong evidence for choosing one over another. The purpose of this work is to compare and understand the effectiveness of five depth sonifications in order to assist the design process of future visual to auditory systems for blind people which rely on depth sensors.
The frequency, amplitude and reverberation of the sound as well as the repetition rate of short high-pitched sounds and the signal-to-noise ratio of a mixture between pure sound and noise are studied. 
We conducted positioning experiments with twenty-eight sighted blindfolded participants. Stage 1 incorporates learning phases followed by depth estimation tasks. Stage 2 adds the additional challenge of azimuth estimation to the first stage's protocol. Stage 3 tests learning retention by incorporating a 10-minute break before re-testing depth estimation. 
The best depth estimates in stage 1 were obtained with the sound frequency and the repetition rate of beeps.
In stage 2, the beep repetition rate yielded the best depth estimation and no significant difference was observed for the azimuth estimation. Results of stage 3 showed that the beep repetition rate was the easiest sonification to memorize. Based on statistical analysis of the results, we discuss the effectiveness of each sonification and compare with other studies that encode depth into sounds. 
Finally we provide recommendations for the design of depth encoding. 



\end{abstract}


%
\IEEEpeerreviewmaketitle

%
%
%
%

\section{Introduction}
\IEEEPARstart{S}{onification} is the use of non-verbal sound to convey information~\cite{kramer1994auditory}. Common examples 
include audio alerts in cars, computers, mobile phones, 
or the sound produced by traffic lights for blind people. 


Sonification of visual information~\footnote{information that is usually captured by vision} can provide essential assistance 
with the audio modality~\cite{edwards2011auditory}. 
Sensory substitution of vision by audition (SSVA) systems for the blind are one straightforward application of such sonification. 
Early SSVA devices encode 2D camera images into sounds and have shown potential for visual rehabilitation~\cite{auvray2007learning,proulx2008seeing,Brown2011,Renier2010,durette2008,hanneton2010}.
A new generation of SSVA devices takes advantage of the rapid development of 3D sensors to explicitly sonify depth~\cite{stoll2015navigating,Brock2013,Massiceti2018,neugebauer2020,yang2016expanding,yang2018long,milios2003sonification,aladren2016navigation,Skulimowski2018,Li2020,Kayukawa2019,Bai2019,Ribeiro2012,hamilton2021soundsight}. 

A variety of depth sonifications are used in devices that rely on 3D sensors. 
Some systems encode depth into sound amplitude to sonify 3D images from a~\textit{Kinect} camera~\cite{stoll2015navigating,Brock2013}, locate object within virtual environments~\cite{Massiceti2018,neugebauer2020} or indicate the length of the traversable area in front of users~\cite{yang2016expanding,yang2018long}. 
Other devices encode depth into pitch for 3D space perception~\cite{milios2003sonification}, sightless navigation~\cite{aladren2016navigation,Skulimowski2018} or hazard detection~\cite{Li2020}. Others use the repetition rate of beeps~\footnote{short, high-pitched periodic sounds} to encode the distance of pedestrians in front of users~\cite{Kayukawa2019} or to warn them of nearby obstacles~\cite{Bai2019}.
Finally, some systems use reverberation to represent the depth of objects~\cite{Ribeiro2012} or to inform users about the size of a room~\cite{hamilton2021soundsight}.  


To date, there is no standardized depth sonification for SSVA 
systems based on an extensive study. 
Yet, several works underline the importance of choosing effective vision to audition mappings for designing such systems~\cite{kristjansson2016designing,Tsiros2017,hamilton2016cross}. In~\cite{kristjansson2016designing}, the authors claim that more research is needed to answer the question ``what sort of stimulation is most effective for conveying particular information ?''. In~\cite{Tsiros2017}, the authors argue that the design of human machine interface systems could greatly benefit from the study of natural cross modal correspondences (CMCs)~\footnote{CMCs are defined as a ``tendency for a sensory feature in one modality to be associated with a sensory feature in another sensory modality''~\cite{spence2011crossmodal}.}. In~\cite{hamilton2016cross} the authors showed that the use of CMCs as a basis for designing color to sound association rules improved performance in memory and recognition tasks. 
 
In this work, we compare the effectiveness of five different depth sonifications for SSVA devices that rely on 3D sensors. 
In such systems, 3D sensors can usually capture depth up to 8 meters. 
We developed a system and protocol to precisely evaluate the accuracy that can give different depth sonifications in a depth range of 1 meter in front of the user.
We discuss in section~\ref{sec:limitation} whether our results could be extended to larger depth ranges. 

To our knowledge, sonification comparison works 
of short range depth (i.e. with a depth that is within the range of 3D sensors used in SSVA devices) were always conducted with sighted participants and are reported below.

Parseihian~\textit{et al.}~\cite{Parseihian2012} found that repetition rate of beeps and pitch were more accurate than reverberation to encode the distance between the position of the hand and a reference point. 
Later, Parseihian~\textit{et al.}~\cite{Parseihian2016} evaluated the potential of 4 sound parameters (pitch, rhythm, timbre and loudness) for a one dimensional guidance task. 
They found that performance was dependent on both the sonification strategy and the given instructions. 
Bazilinsky~\textit{et al.}~\cite{Bazilinskyy2016} investigated, with 29 sighted individuals, 3 sound parameters (pitch, amplitude and repetition rate of beeps) to encode distance between the cursor and a reference point on a computer screen. 
The authors did not find significant differences in the estimation accuracy between the 3 sonifications. 


The aforementioned sonification comparisons were performed with sighted non-blindfolded individuals. 
Participants therefore had visual feedback combined with audio during the learning phase as well as during the position estimation and guidance tasks. It is reasonable to hypothesize that behavioral and motor control differences exist when people cannot rely on vision to estimate the audio encoded depth. 
Therefore, in the context of sensory substitution for blind individuals, we compare sonifications without visual feedback. In addition, we conducted experiments 
with a physical object to imitate a real life situation in which a blind person wants to understand the position of an object. The importance of using a physical object in the sensory substitution context has been highlighted by Auvray~\textit{et al.}~\cite{auvray2005there}. They 
showed that the distal attribution~\footnote{The distal attribution is the ability to create a mental representation of a distant object from sensory stimulation} in the sensory substitution process relies on the ability to manipulate the physical object that is encoded with sounds. 
Finally, conducting experiments in the real world rather than in a virtual environment allows to remove  performance biases due to different skill levels of participants with controllers such as a game pad or a mouse. 

Our protocol comprises three experimental stages that are described in section~\ref{sec:protocol}. 
Experiments were conducted with 28 sighted blindfolded participants. 
Participant's position estimation accuracy was measured for five different sonifications. 
At the end of the experiment, participants completed a short questionnaire to give qualitative feedback on the sonifications. We also conducted an online experiment to estimate Just Noticeable Differences (JND), which we report on in the discussion section. 
 

Of the five investigated sonifications, two are based on natural depth cues used to estimate the distance of a sound source: the amplitude which is commonly used in sonification
systems, and the reverberation (as in the work of  Ribeiro~\textit{et al.}~\cite{Ribeiro2012}). 
These may provide an intuitive way to encode depth. However, humans tend to naturally underestimate and overestimate far and near sound sources~\cite{zahorik2005auditory,kolarik2016auditory}.
We therefore tested 3 other sonifications based on artificial cues that could potentially overcome these natural biases. 
One is a mix of a pure tone and a white noise. The amplitude ratio of the two signals is used to encode depth. Plazak et al.~\cite{Plazak2017} showed that this sonification can be used to assist neurosurgeons to encode distance between a surgical probe and an anatomical region. 
To our knowledge, this is the first time that such a sonification is investigated in a sensory substitution context. Finally, we investigated two other common depth sonifications used in vision to audition devices that rely on depth: the frequency, and the repetition rate of beeps. It is hypothesized that the 5 evaluated sonifications give significantly different depth estimation accuracy and user preferences.

\section{Depth sonification system}
\label{sec:distToSound}

\begin{figure}[htb] 
   \centering
   \includegraphics[width=0.29\textwidth]{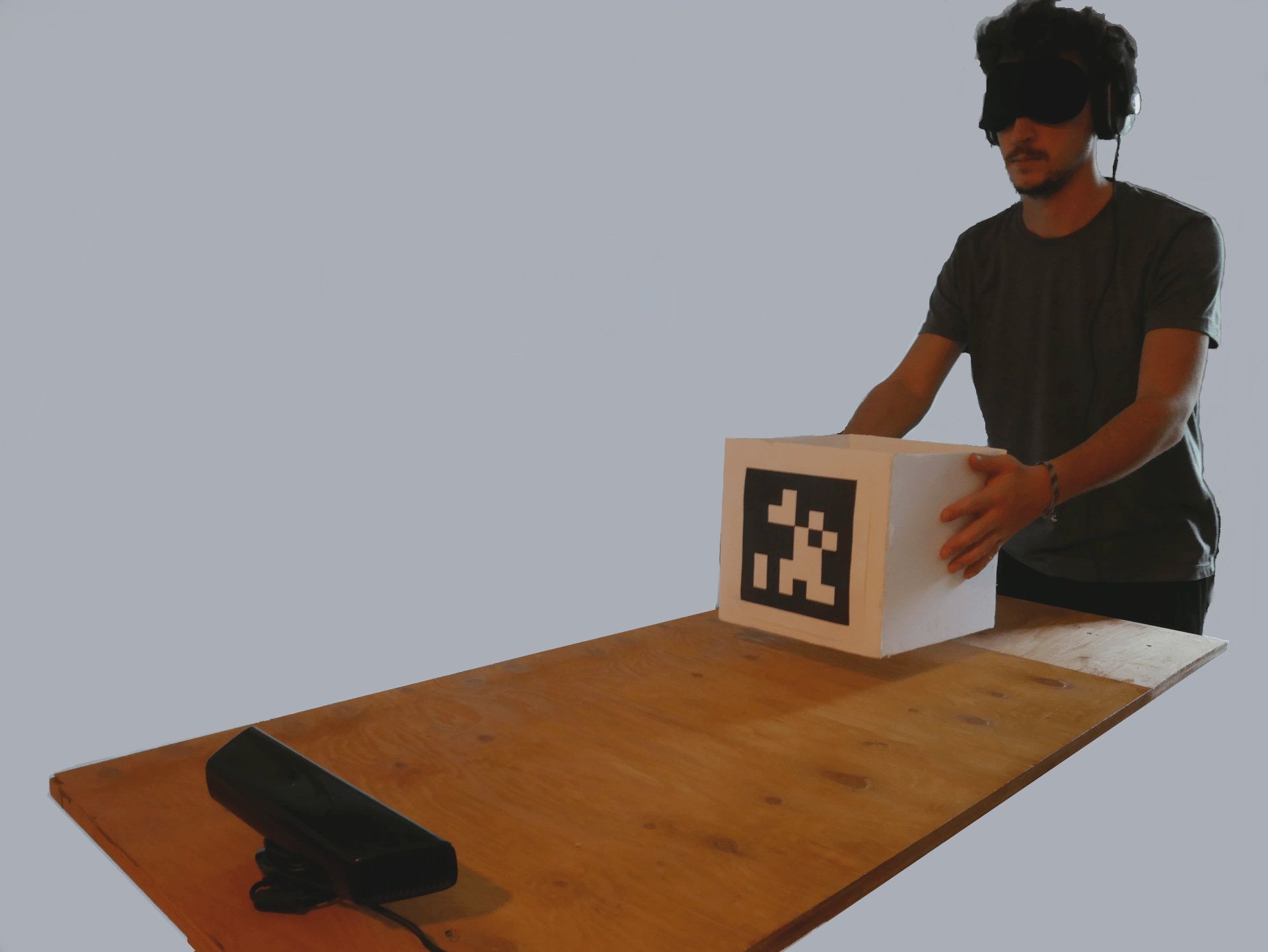} 
   \caption{Experimental setup. Participants are blindfolded and standing in front of the table. The setup comprises the box with the April Tag, a camera that captures the 3D position of the box, and a computer that runs the position tracking and the sonification algorithm. 
   }
      \label{fig:expSetupPhoto}   
\end{figure}

The system consists of a styrofoam box along with a camera and a computer to capture and sonify the position of the box (Fig.~\ref{fig:expSetupPhoto}). 
The audio feedback is delivered through a pair of~\textit{Audio-technica} headphones (model ATH-M50X).  

\begin{figure}[htb] 
   \centering
   \includegraphics[width=0.45\textwidth]{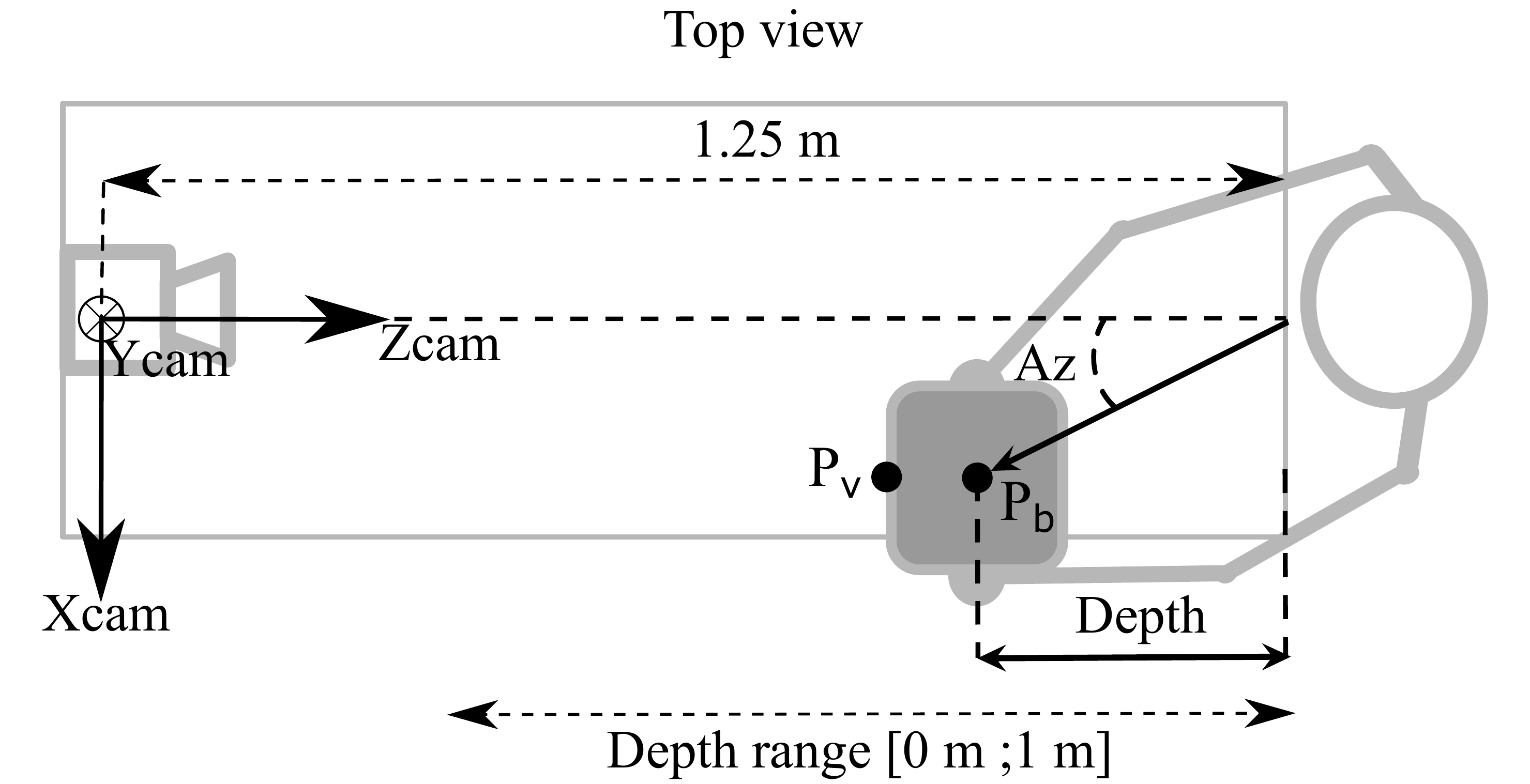} 
   \caption{Schematic top view of the experimental setup. The camera's coordinate system $C_{cam}$ comprises the horizontal $Zcam$ axis pointing towards the participant, the vertical $Ycam$ axis pointing down and the $Xcam$ axis pointing toward the left of the participant. $P_v$ and $P_b$ are the positions of the visual marker and the box in $C_{cam}$. 
   The depth is assumed to be $0~m$ when the box is on the right edge of the table. 
   $Az$ is the azimuth angle between the $Zcam$ axis and the position of the box  $P_b$, with the origin of the angle being the midpoint of the table on the participant's side. 
   }
      \label{fig:schemaSetupExp}   
\end{figure}

 The system has two modes: one for the learning and one for positioning tasks. During the learning, the position of the styrofoam box is tracked and sonified in real-time while participants are moving the box. For the positioning task, an 
 audio encoded position is played and participants have to move the box to the perceived location. The system encodes either the depth only or the azimuth and the depth (defined in Fig.~\ref{fig:schemaSetupExp}).

The camera is the RGB sensor of a~\textit{XBOX Kinect 360} (model 1414).  
It would be possible to use the depth sensor of the~\textit{Kinect}, but we designed our experiments to be reproducible with a simple RGB camera. The position tracking of the box is implemented in~\textit{Python}. Sound is synthesized with the~\textit{Supercollider}~\cite{mccartney2002rethinking} audio programming environment. 
  
 

 \subsubsection{Box position tracking}
 \label{sec:posTracking}
 The position of the box is estimated in real time thanks to a visual marker pasted on the box (Fig.~\ref{fig:expSetupPhoto}) that is detected by the~\textit{AprilTag} algorithm~\cite{olson2011tags}. 
~\textit{AprilTag} estimates 3D positions of visual markers relative to a camera. We detail below the computation of the depth and azimuth (Fig.~\ref{fig:schemaSetupExp}).

We denote $P_b(x_b,y_b,z_b)$ the position of the box and $P_v(x_v,y_v,z_v)$ the position of the visual marker in the camera's coordinate system $C_{cam}=(Xcam,Ycam,Zcam)$ (Fig.~\ref{fig:schemaSetupExp}). 
~\textit{AprilTag} gives the position of the visual marker $P_v(x_v,y_v,z_v)$ in $C_{cam}$. 
We first compute the position of the box from the position of the visual marker: $(x_b,y_b,z_b) = (x_v,y_v,z_v + s_b/2)$, with $s_b$ being the length of the box edges, equal here to $28$ cm.
 We then estimate the depth in centimeters as $depth = 125 - z_b$.  
 We also compute the azimuth (in degree) with:
$$Az = \begin{cases} \arctan(-x_{b}/depth) & \textrm{if} \quad{} depth \neq 0\\ -90 & \textrm{if} \quad{} depth = 0, x_{b} >0 \\ 90& \textrm{if} \quad{} depth = 0,  x_{b} < 0\end{cases}$$

 


\subsubsection{Sonification design}
\label{sec:sonifDesign}
\begin{table*}[h!]
     \centering
        \captionof{table}{
        Sound parameter ranges and equations for each sonification. The encoding parameter varies linearly with depth. The range column ($[P_{d=0m}, P_{d=1m}]$) shows the sound parameter values for a depth of 0 and 1 meter. 
       For~\textit{Freq}, the signal amplitude $A(M(d))$ depends on the frequency of the sound signal and is set with the ``equal loudness curve'' of the standard~\textit{ISO 226:2003}~\cite{ISOEqualLoudness}. 
       For~\textit{BRR} (Beep Repetition Rate) and~\textit{Reverb}, $ExpEnv$* is an exponential decay envelope defined on $t \in [0,1/\tau]$, equal to  $exp(-39\cdot t$). $\tau=1$ second for~\textit{Reverb} and $\tau$ depends on the depth $d$ for~\textit{Beep}. The envelope $ExpEnv$ reduces the amplitude of the sound signal by 98\% after $0.1$ second. The envelope is repeated every $1/\tau(d)$ second for~\textit{Beep} and every 1 second ($\tau=1$ second) for~\textit{Reverb}. For~\textit{Reverb}, we the use~\textit{Freeverb} filter, which is a public implementation of Schroeder's reverberators~\cite{schroeder1961}. 
       For~\textit{SNR}, $N(t)$ is a white noise. 
        }
        
\renewcommand{\arraystretch}{1.5} 
 \begin{tabular}{cccc}
    \toprule
                  Sonification  & Encoding parameter &\makecell{$[P_{d=0m}, P_{d=1m}]$} &Sound signal s(t): \\ \midrule
    		 \rowcolor[HTML]{EFEFEF} 
~\textit{Freq} & $M(d)$: Midi note number &  $[107, 48]$& $A(M(d))\cdot sin(2\pi \cdot (440 \cdot 2^{(M(d)-69)/12)})\cdot t)$ \\ 
~\textit{Amp}  & $A(d)$: Amplitude (DB) &$[0,-40]$& $10^{-A(d)/20} \cdot sin(2\pi \cdot 500 \cdot t)$ \\
		\rowcolor[HTML]{EFEFEF} 
~\textit{Reverb}  & $RT(d)$: Reverberation time &$[0.05,0.95]$& $Freeverb_{RT}[(\sum_{k=0}^{\infty }ExpEnv\text{*}(t-k/1))\cdot sin(2\pi \cdot 1200 \cdot t)]$ \\
~\textit{BRR} & $\tau(d)$: Repetition rate (Hz) &$[10 ,1]$& $ (\sum_{k=0}^{\infty }ExpEnv(t-k/\tau))\cdot sin(2\pi \cdot 1200 \cdot t) $ \\
       	\rowcolor[HTML]{EFEFEF} 
~\textit{SNR}  & $R(d)$: Signal-Noise ratio& $[20;0.05]$& $R(d)/(1+R(d)) \cdot sin(2\pi \cdot 500 \cdot t)+1/(1+R(d))\cdot N(t)$ \\ \bottomrule
  \end{tabular}
    \label{tab:parameters}
\end{table*}	

Each depth sonification algorithm has one sound parameter $P(d)$ that 
is linearly mapped to the depth $d$ with the following equation:
\begin{equation}
\label{eq:linearInterp}
P(d) = d \times (P_{d=1m}-P_{d=0m}) +P_{d=0m}
\end{equation}
with $P_{d=0m}$ and $P_ {d=1m}$ (described in Table~\ref{tab:parameters}) being the values of sound parameters used to encode a depth $d$ of 0 and 1 meter, respectively. Outside of the $[0;1]$ meter depth range, sound parameters are not changed. 


Table~\ref{tab:parameters} gives the sound parameter ranges and equations for each sonification. We provide additional details below:  
\begin{itemize}[leftmargin=0cm,itemindent=.5cm,labelwidth=.2cm,labelsep=0cm,align=left]
\item~\textit{Freq}: A pure tone with frequency varying according to the western chromatic musical scale. 
The sound frequency is therefore equally distinguishable within all the depth range. 
Small and large depths are respectively encoded into high and low pitch notes. The amplitude is weighted according the ``equal loudness curve'' of the standard~\textit{ISO 226:2003}~\cite{ISOEqualLoudness} to ensure that the perception of loudness is equal throughout the frequency range. 
 \item~\textit{Amp}: The amplitude of a pure tone ($500 Hz$) according to a decibel scale. 
 Small and large depths are respectively encoded into high and low sound amplitude.
 \item~\textit{Reverb}: the reverberation time of beeps. 
Here, beeps are high pitched pure tones 
 that lasts $200~ms$ and that are repeated at a fixed rate of one beep per second. Beeps are reverberated with Schroeder reverberators~\cite{schroeder1961} implemented by the synthesizer~\textit{FreeVerb} of~\textit{Supercollider}.~\textit{FreeVerb} consists of 8 low-pass feedback comb filters in parallel followed by 4 all-pass feedback filters in series~\cite{JuliusReverb}. 
 The reverberation time is defined as the time for the low frequencies (below 1500Hz) of the sound to be reduced by 60 dB. 
 Small and large depths are encoded respectively into short and long reverberation time. 
 \item~\textit{Beep Repetition Rate (BRR)~\footnote{We use the same notation as proposed by~\cite{Bazilinskyy2016}}}: the repetition rate of beeps (high pitch tone of $200~ms$). 
  Small and large depths are encoded into fast and slow repetition rates, respectively. 
 \item~\textit{SNR}: a pure tone ($1200~Hz$) mixed with white noise. 
 The varying parameter is the ratio of the amplitude of the pure tone over the amplitude of the white noise. Small and large depths are encoded respectively into high amplitude of the pure tone (low amplitude of the noise) and high amplitude of the noise (low amplitude of the pure tone). 
\end{itemize}

When sonified, the azimuth angle (Fig.~\ref{fig:schemaSetupExp})  is linearly mapped to stereo panning. The stereo panning is the distribution of the volume in the left and right channels of the headphones. Azimuths of $-90\degree$ and $90\degree$~are encoded with the signal being entirely played in the left ear and the right ear, respectively. 

\section{Methods}
\label{sec:method}
\subsection{Participants}
Twenty-eight voluntary subjects were recruited from the~\textit{Université de Sherbrooke} (11 women and 17 men; mean age: $27~\pm 5.8$ years). All were naive regarding the purpose of the experiment and none
of the subjects reported any hearing losses.
None of them were relatives or friends of the authors. Each individual received the same amount of financial compensation. 


\subsection{Design}
We performed a within-subjects study where each participant performed the tasks with the 5 different sonifications. 
Independent variables are the depth sonification type and the 3 experimental stages (describe in section~\ref{sec:protocol}). Dependent variables are the depth estimation accuracy and participants preferences. 

\subsection{Procedure}

\label{sec:protocol}
Individuals were standing in front of a table and listening with headphones (Fig.~\ref{fig:expSetupPhoto}). 
Participants were first asked to calibrate the volume of the loudest sound in the experiment by increasing it while remaining at a comfortable level.
The protocol was divided into three experimental stages~\footnote{The 3 stages were approved by the ethical committee from ``lettres et sciences humaines'' faculty from Universit\'{e} de Sherbrooke under reference number 2014-85/Rouat.}, each completed with the 5 sonifications. 
Before the beginning of each stage, participants were blindfolded.

     

\begin{enumerate}[leftmargin=0cm,itemindent=.5cm,labelwidth=.2cm,labelsep=0cm,align=left]
\item \textit{  Stage 1 - Depth estimation with learning tasks}
:  
the stage comprised 3 sequences of [1 learning task followed by 5 positioning tasks] (total of 3 learning and 15 positioning tasks). During the learning task, participants were asked to actively explore the space by moving the box and to try to understand
how the sound changed depending on its position. There was no time limit for this task. For the positioning task, a random target position (either a depth or a depth and an azimuth for stage 2) was generated from an uniform distribution. The generated target was constrained to the area within the field of view of the camera in front of the participant. 
A two second sound encoding the target position was then played and participants had to place the box at the perceived location. 

The purpose was to quantify how well participants estimated the depth and how the learning tasks impacted the positioning accuracy.
Only the depth was encoded into sounds.
\item \textit{    Stage 2 - Depth and azimuth estimation}
: the purpose was to investigate the simultaneous perception of the encoded azimuth and depth. 
The stage comprised one learning task followed by 5 positioning tasks. Learning and positioning tasks were the same as those described above for stage 1, but with both depth and azimuth encoded into sounds. 
\item  \textit{    Stage 3 - Depth estimation after a 10 minutes break}:  the purpose was to evaluate how well the different sonifications could be remembered in the short term. Participants therefore took a 10 minutes break before performing the stage and there was no new learning task. 
Then they had to complete 5 positioning tasks, which were the same as in stage 1. 
\end{enumerate}

Participants could rest and remove their blindfolds whenever they needed during experiments. We did not provide feedback on the accuracy of their estimation. 
Following the experiments, they answered a questionnaire comprising 4 questions to give qualitative feedback on the sonifications.

\subsection{Graphical representation and statistical analysis}
\label{sec:stats}
Distribution of depth and azimuth errors are displayed with Boxplots~\cite{rice2006mathematicalBoxplot}. Horizontal edges of the box show first and third quartiles and ends of the vertical lines represent minima and maxima. Medians and averages are shown with middle line in the box and  black filled circles, respectively. 
Outliers~\footnote{Data points that are more than 1.5 Interquartile Range (IQR) away from the first or third quartile} are shown with with empty black circles. 

We define the chance level as the expected 
error if participants had randomly positioned the box without the help of sonifications. 
The computation of the chance level is shown in Appendix section~\ref{annex:guessChances} and is equal to $1/3$ of the range (i.e. $100/3=33~cm$ for the depth and $180/3=60\degree$ for the azimuth).

To analyze the effect of sonifications on participant's performance, we use different versions of the ANalysis Of Variance (ANOVA)~\cite{kennyANOVAandPostHoc}. An ANOVA determines whether the means of two or more distributions are different, by comparing inter- and intra-group variances. For each ANOVA, we give the F statistic~\footnote{The F statistic represents the ratio between the inter- and the intra- group variance. F is computed from the Fisher distribution and the degree of freedom of the inter- and intra- group.} and the p-value~\footnote{The p-value is computed from the F statistic and represent the probability of obtaining the observed means by chance}. The significance level for the ANOVAs is set with the p-value at $p<0.05$. 

A two-factors~\footnote{Several factors ANOVAs are used for analyzing the effect of several independent variables on one outcome variable} repeated measure~\footnote{The repeated measure ANOVA is used when data are collected from the same individuals under different conditions or at different times} ANOVA is used to analyze the effect of learning and sonifications on depth errors (section~\ref{sec:stage1}). One-factor repeated measure ANOVAs are used to study the effect of  sonifications on depth (section~\ref{sec:stage2},\ref{sec:stage3}) and azimuth  (section~\ref{sec:stage2}) errors. When ANOVAs are significant (i.e. $p<0.05$), multiple pairwise posthoc~\cite{kennyPostHoc} t-tests are performed to determine which sonification pairs yield significant different errors. The significance level~\footnote{Significance levels of the multiple pairwise comparisons were adjusted with the Holm method.} of the pairwise comparisons are shown with gray asterisks above boxplots as follows: * for $\left\{p<0.05\right\}$, ** for $\left\{p<0.01\right\}$ and *** for $\left\{p<0.001\right\}$.

Simple linear regressions are computed to analyze to model the estimated depths or azimuths and the actual target depths (section~\ref{sec:stage1}) or azimuths (section~\ref{sec:stage1}) generated by the system. For each linear regression, we give the goodness of fit $R^2$~\footnote{ The goodness of fit $R^2$ is the percentage of the variance of the dependent variable explained by the linear model} and the regression equation. On each linear regression figure (Figs.~\ref{fig:linReg} and~\ref{fig:linRegAngleAngle}), the dashed line is the identity function (i.e. when the target depth is equal to the depth estimate). The plain black line shows the linear model. The shaded area represents the 95\% confidence interval of the linear model.


\section{Results}
Results are reported for 28 blindfolded sighted individuals. Of the 28 participants, 14 completed the experimental stages 1 and 2 and 14 the stages 1 and 3. One participant felt sick during the experiment and completed only stage 1 and the questionnaire. Results of the first, second and third experimental stages are therefore reported for 28, 13 and 14 individuals. 

\subsection{stage 1, depth estimation with learning tasks}

\label{sec:stage1}
In the rest of the paper, positioning errors refer to absolute positioning errors. 

 We consider a sonification to have failed to encode depth for individuals when their average depth error over the 15 positioning tasks was above the chance level ($33~cm$). 
During this stage, one participant had an average error above the chance level with~\textit{Amp} ($40.3~cm$) and~\textit{Reverb} ($38.1~cm$), while a second had an error level above chance with~\textit{Reverb} ($35.0~cm$). 

\begin{figure}[htb] 
   \centering
   \includegraphics[width=0.46\textwidth]{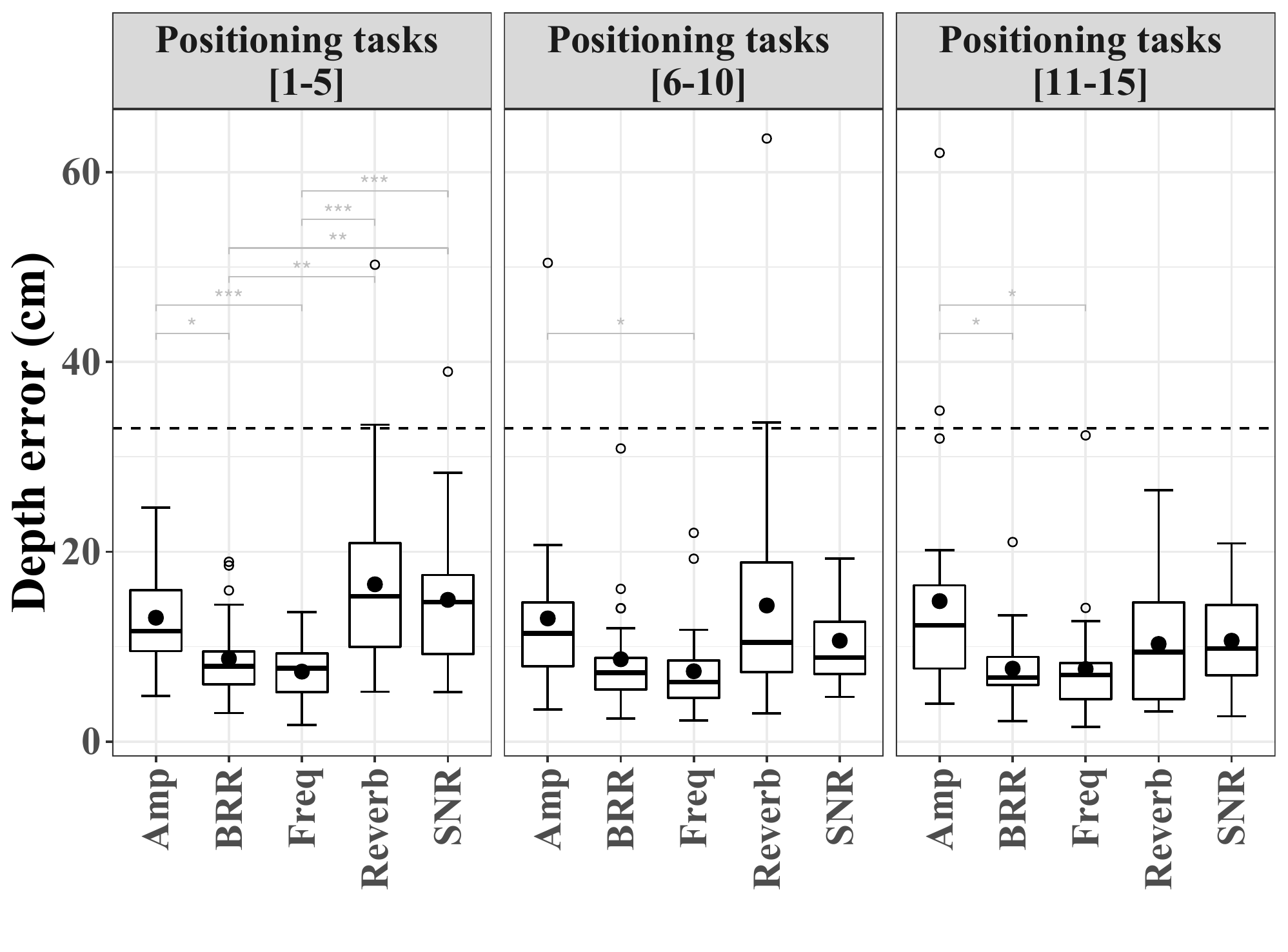} 
   \caption{Distribution of participants' average depth errors for each sonification after the first (positioning tasks [1-5]), second  (positioning tasks [6-10]) and third learning task (positioning tasks [11-15]). See section~\ref{sec:stats} for a detailed description of the boxplots. 
   }
      \label{fig:errVsSound1}    
\end{figure}

Figure~\ref{fig:errVsSound1} shows the distribution of 
participants' average depth errors with each sonification after each learning task. 

The two-factor repeated measure ANalysis Of Variance (ANOVA) showed a statistically significant interaction between the effect of the sonification and learning tasks on depth errors ($F_{4.87,126.58}~\footnote{Non integer degrees of freedom for the F statistic are due to the Greenhouse Geisser correction applied to adjust the non-equality of variances, which is a necessary assumption to conduct a repeated-measure ANOVA} = 9.58,~p = \num{0.015}$). Simple main effect analysis showed that sonifications had a significant effect on depth errors after the first (positioning tasks [1-5]) ($F_{2.81,73.0}= 11.6,~p = \num{4.41e-06}$), second (positioning tasks [6-10]) ($F_{2.34,60.9}= 4.53,~p = \num{0.011}$) and third (positioning tasks [11-15]) learning task ($F_{2.35,61.0}= 6.04,~p = \num{0.003}$). 
Pairwise comparisons further indicated that (i)~\textit{Freq} and~\textit{BRR} (Beep Repetition Rate) were significantly better than the 3 other sonifications after learning task 1, (ii)~\textit{Freq} was significantly better than~\textit{Amp} after learning task 2 and (iii)~\textit{Freq} and~\textit{BRR} were significantly better than~\textit{Amp} after learning task 3 (Fig.~\ref{fig:errVsSound1}).~\textit{Freq} and~\textit{BRR} are the most effective 2).~\textit{Reverb} and~\textit{SNR} are initially not effective but, with practice, results improve.~\textit{Amp} is not effective, even after practice. 
Smaller Interquartile ranges of depth errors obtained with~\textit{Freq} and~\textit{BRR} shows more consistency between participants' performances with these 2 sonifications. 

Simple main effect analysis also showed that learning tasks had a significant effect on depth errors with~\textit{Reverb} ($F_{2,52}= 5.83,~p = \num{0.005}$) and~\textit{SNR} ($F_{1.61,41.8}= 6.51,~p = \num{0.006}$). Pairwise comparisons further showed that (i)~\textit{Reverb} led to significantly better accuracy after the learning task 3 than during after the learning task 1 ($p=\num{0.006}$) and (ii)~\textit{SNR} led to significantly better accuracy after learning task 2 ($p=\num{0.017}$) and 3 ($p=\num{0.026}$) than after learning task 1. This supports the idea that practice with~\textit{Reverb} and~\textit{SNR} allowed participants to increase their positioning accuracy.


 
   
%

\begin{figure}[htb] 
   \centering
   \includegraphics[width=0.47\textwidth]{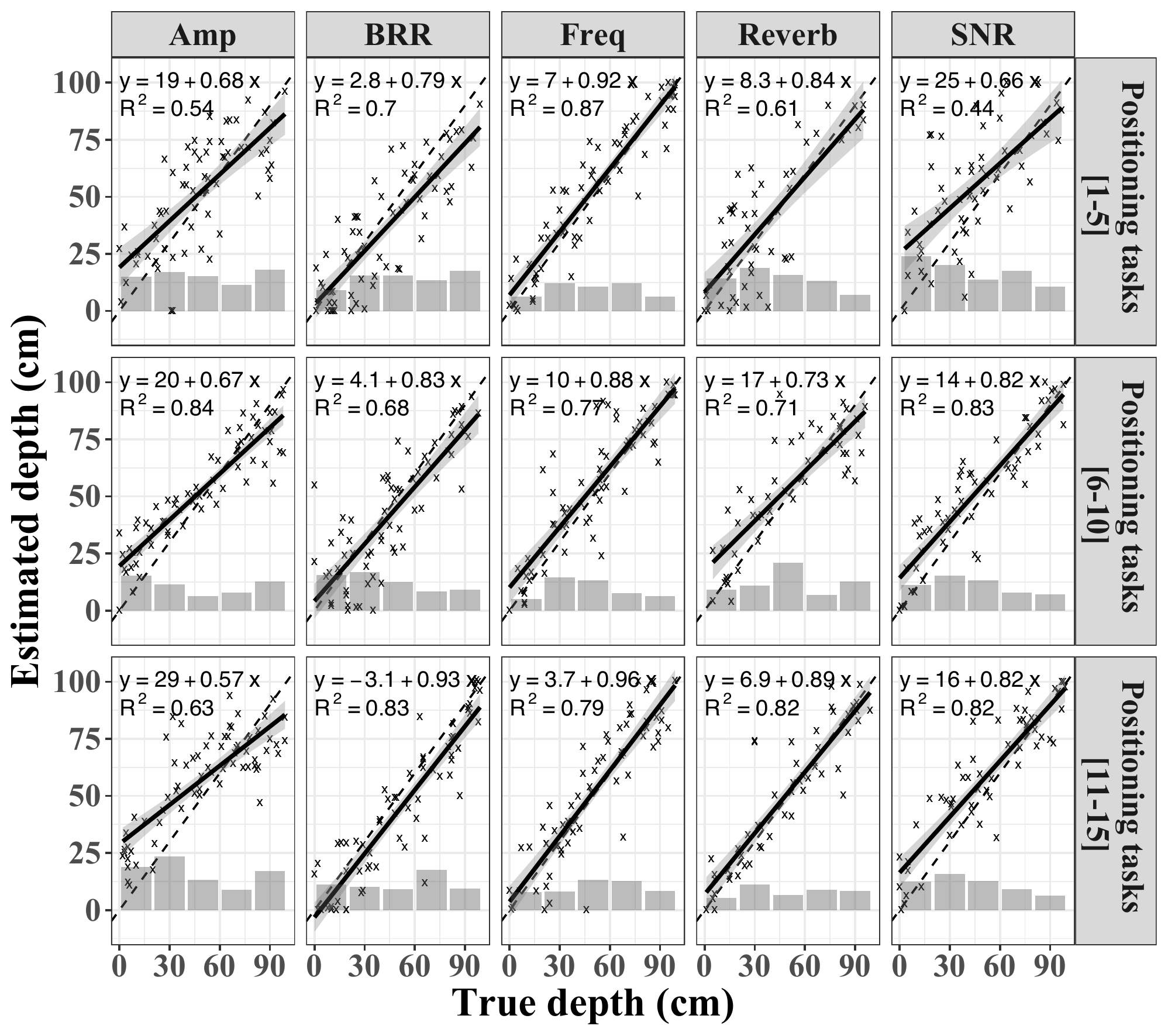} 
   \caption{Linear regression to model participants' depth estimates as a function of true target depths. Gray barplots show the average errors as a function of the true target depth. We did not identify a clear pattern of the average error distribution with respect to the actual target depth. 
   }
      \label{fig:linReg}   
\end{figure}

Linear regressions were estimated to model perceived depths as a function of target depths (Fig.~\ref{fig:linReg}). We define the depth estimation bias as the difference between the regression curve and the identity function (showed with dashed lines in Fig.~\ref{fig:linReg}). 
To compute the regressions, we removed data when a sonification failed to encode depth for one participant (i.e. when the average depth error was above the chance level). 

Significant regression was found for each sonification and phase ($p<\num{2.2e-16}$), suggesting that estimates linearly change with target depths.
~\textit{Amp} resulted in the largest estimation biases with a clear tendency to overestimate small depths and underestimate large depths. 
~\textit{Freq} gave the smallest estimation biases over the 3 phases.~\textit{BRR, Reverb}~\textit{and SNR} gave similar estimation biases. Except with~\textit{Amp}, estimation biases were smaller in phase 3 than in phases 1 and 2.    
 


\subsection{stage 2, depth and azimuth simultaneous estimation}
\label{sec:stage2}

\begin{figure}[htb] 
   \centering
   \includegraphics[width=0.45\textwidth]{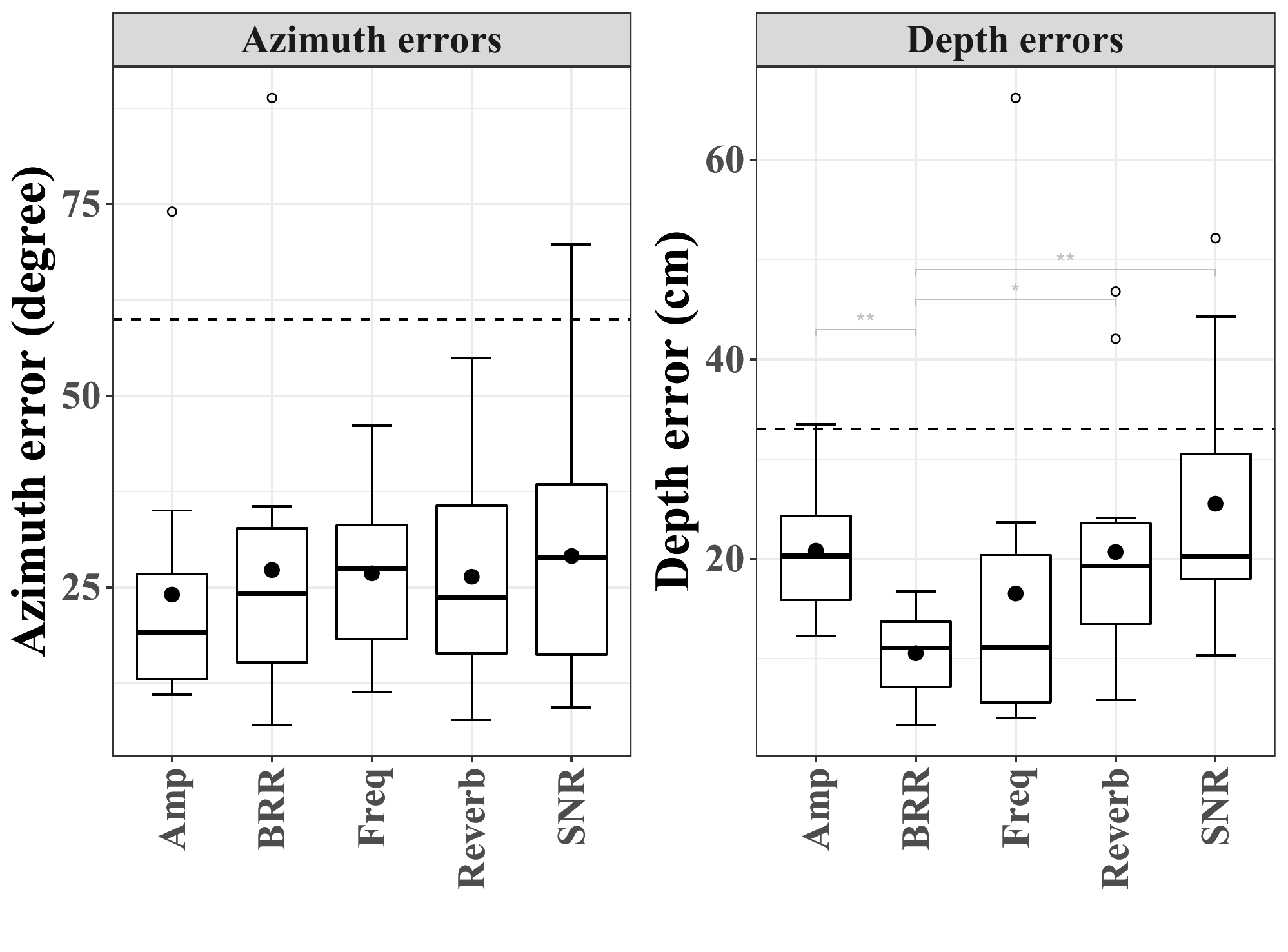} 
   \caption{Distribution of average azimuth (left) and depth (right) errors during the experimental stage (simultaneous depth and azimuth estimation). 
   }
      \label{fig:resultAnglePos}   
\end{figure}

Figure~\ref{fig:resultAnglePos} presents the distribution of 
participants' average depth and azimuth errors. 
Two had their average azimuth error over the chance level. One with~\textit{Amp} and~\textit{SNR} with an average azimuth error of $74\pm13\degree$ and $70\pm35\degree$) 
 and one with~\textit{BRR} (Beep Repetition Rate) with an average azimuth error of $89\pm13\degree$. 
These participants reported to be confused by the encoding of the azimuth angle as they rather expected the $y$ cartesian coordinate (Fig.~\ref{fig:schemaSetupExp}) of the box to be audio encoded. 

Although~\textit{Amp} gave slightly better azimuth accuracy, close averages and large variances suggest that there were no significant differences between the sonifications (Fig.~\ref{fig:resultAnglePos}). 
This was confirmed by a repeated measure ANOVA ($F_{4,48} =0.243, p=0.913$).
This suggests that the perception of the audio encoded azimuth is independent of the depth sonification. 

Depth errors were significantly higher during this stage 
than during the first stage. The average depth error over all sonifications and all participants was $21.5\pm13.9~cm$ (compared to $12.3\pm11.3~cm$ for the first stage). Five out of thirteen individuals had their mean depth errors over the chance level ($33~cm$) with at least one of the following sonifications:~\textit{Amp, Freq, Reverb} and~\textit{SNR}. The simultaneous estimation of azimuth and depth was more difficult than depth estimation alone. 

A repeated measure ANOVA showed a significant effect of the sonifications on depth errors ($F_{4,48} =4.731, p=\num{0.003}$). Pairwise comparisons 
indicated that~\textit{BRR} gave significantly better accuracy than~\textit{Reverb, SNR} and~\textit{Amp}. 

\begin{figure}[htb] 
   \centering
   \includegraphics[width=0.5\textwidth]{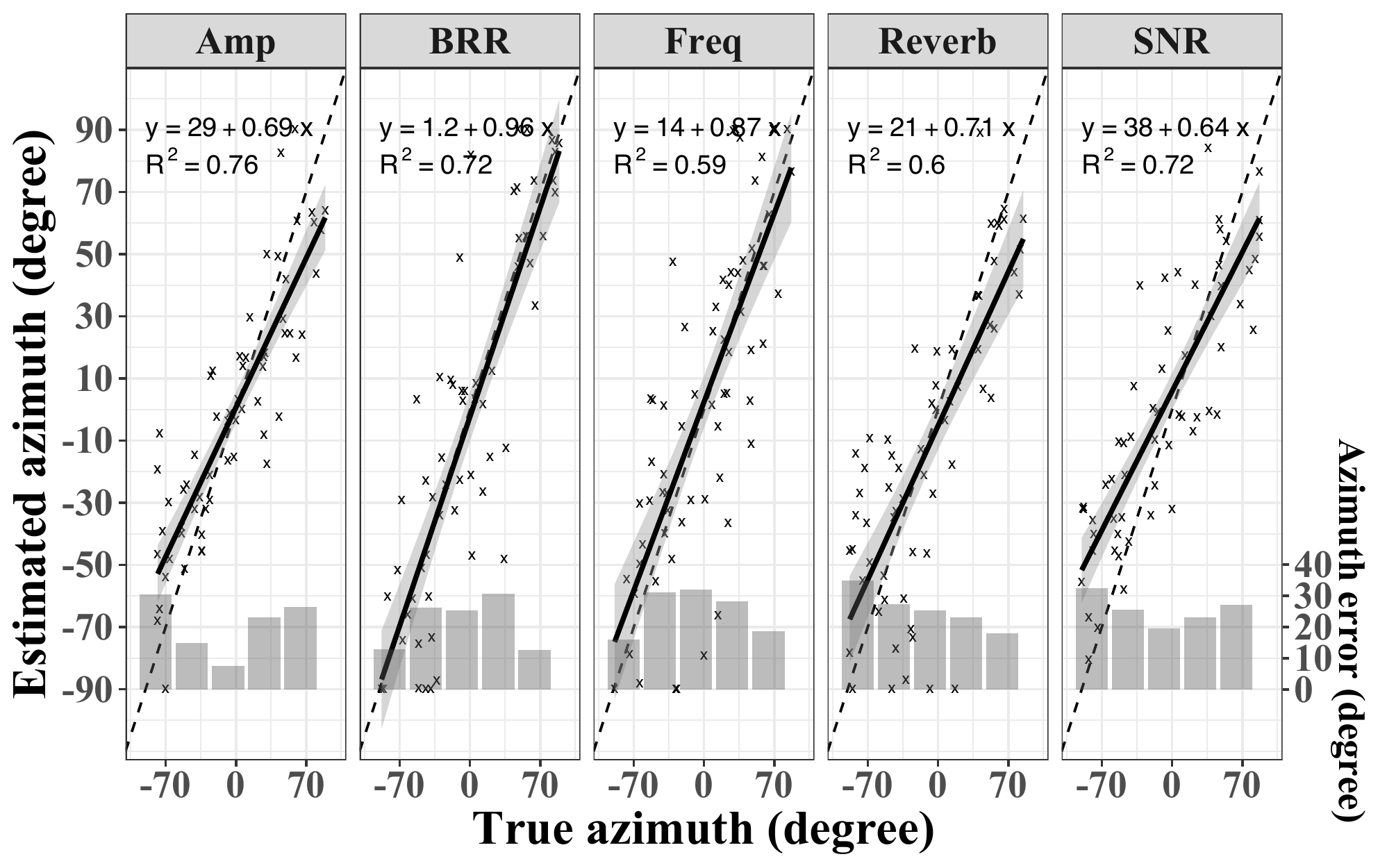} 
   \caption{Linear regression of the participant's azimuth estimation as a function of the true azimuth during the second experimental stage (depth and azimuth estimation). 
The left, center and right position correspond to azimuths of -90\degree, 0\degree~and 90\degree, respectively. Gray barplots show the
average errors as a function of the true target azimuths. }
      \label{fig:linRegAngleAngle}   
\end{figure}

Linear regressions were computed to model the estimated azimuths as a function of the target azimuths (Fig.~\ref{fig:linRegAngleAngle}). We remove participants from the data when their average azimuth error with a sonification was above the chance level. All regressions were significant ($p<\num{2.2e-16}$). 
As for the depth in section~\ref{sec:stage1}, we define azimuth estimation biases as the difference between regression curves and the identity function. 
With~\textit{BRR} and~\textit{Freq}, estimates were almost unbiased. With the other sonifications, participants estimates were biased toward the center (i.e., they underestimated the magnitude of both the left and right azimuth). 

\subsection{stage 3: Depth estimation after a 10 minutes break}
\label{sec:stage3}



\begin{figure}[htb] 
   \centering
   \includegraphics[width=0.38\textwidth]{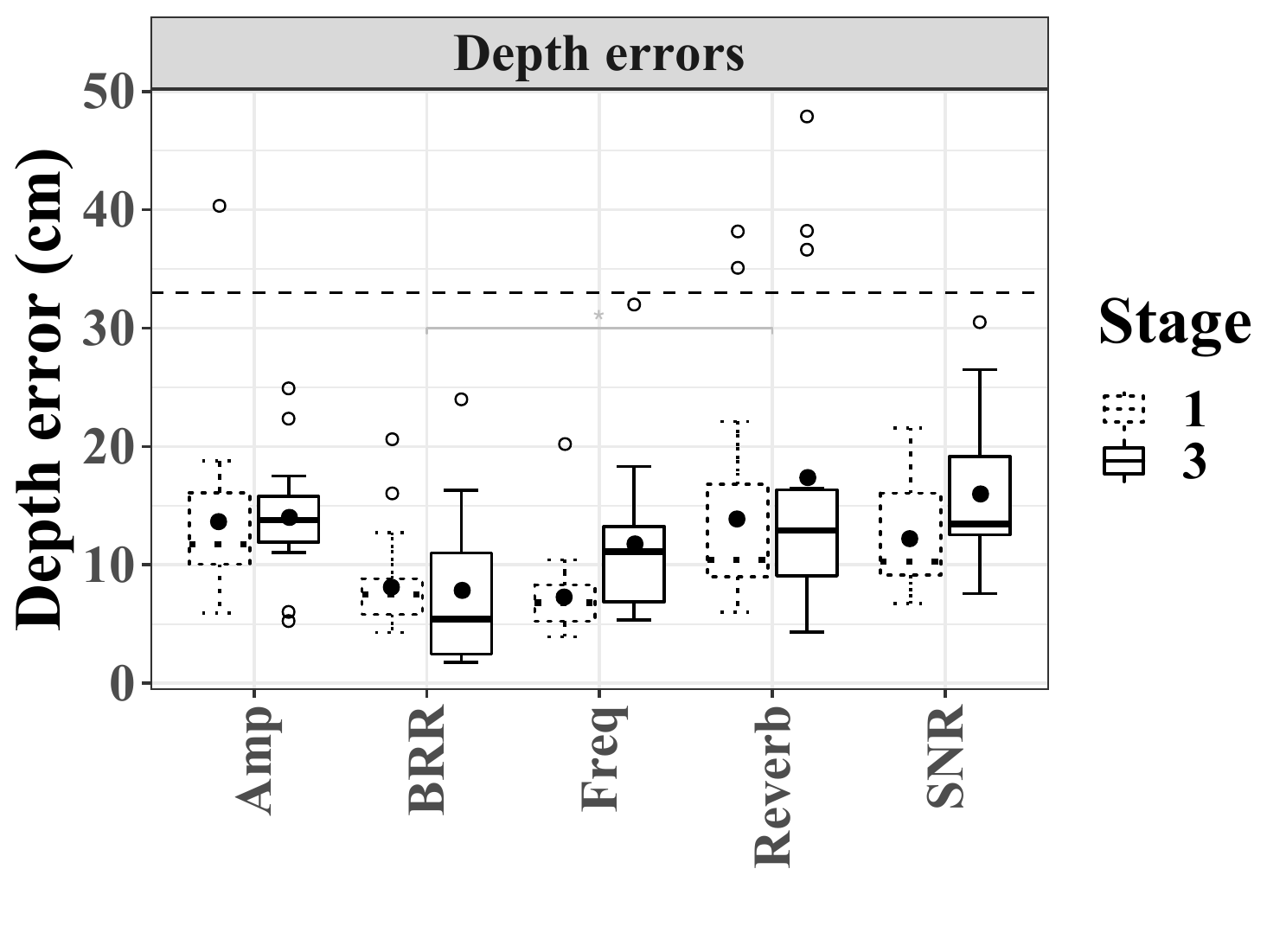} 
   \caption{Distribution of average depth errors during the third experimental stage (depth estimation after a 10 minutes break). For comparison, average results over the 3 phases of stage 1 are shown with dotted lines.  
    }
      \label{fig:resultMemory}   
\end{figure}

Figure~\ref{fig:resultMemory} displays the distribution of participants' average depth errors for stage 3. Distributions are given for the 14 individuals who completed both stages 1 and 3. 
During stage 3, 3 out of 14 participants (21,4\%) had their average depth errors above the chance level ($33~cm$) with~\textit{Reverb} ($36.6\pm15.7~cm$, $38.2\pm16.1~cm$ and $47.9\pm14.1~cm$). This suggests that~\textit{Reverb} is not well remembered. 
A repeated measure ANOVA showed a significant effect of the sonifications on depth errors ($F_{4,52} = 4.452,~p = 0.004$). Pairwise comparisons 
indicated that positioning accuracy with~\textit{BRR} (Beep Repetition Rate) was significantly better than with~\textit{Reverb}. 
Except with~\textit{BRR}, sonifications led to larger positioning errors than in stage 1. With~\textit{BRR},  averages are similar for stages 1 and 3 and the median for stage 3 is even lower than the median for stage 1. 
This suggests that~\textit{BRR} is the easiest sonification to remember.



%

\subsection{Qualitative feedback}
 The questionnaire comprised the 4 following questions:
\begin{itemize} 
 \item (\textit{Easy}) Were sonifications easy to use to estimate depth positions? (rate from 1: ``very easy'' to 5: ``very difficult'')
 \item (\textit{Pleasant}) Were sonifications pleasant to listen to?  (rate from 1: ``very pleasant'' to 5: ``very unpleasant'') 
 \item (\textit{Strategy}) What was your strategy to position the box? (open question)
 \item (\textit{Natural}) What sonification did you find the most natural to encode the depth? (forced choice)
\end{itemize}

\begin{figure}[htb] 
  \centering
  \includegraphics[width=0.42\textwidth]{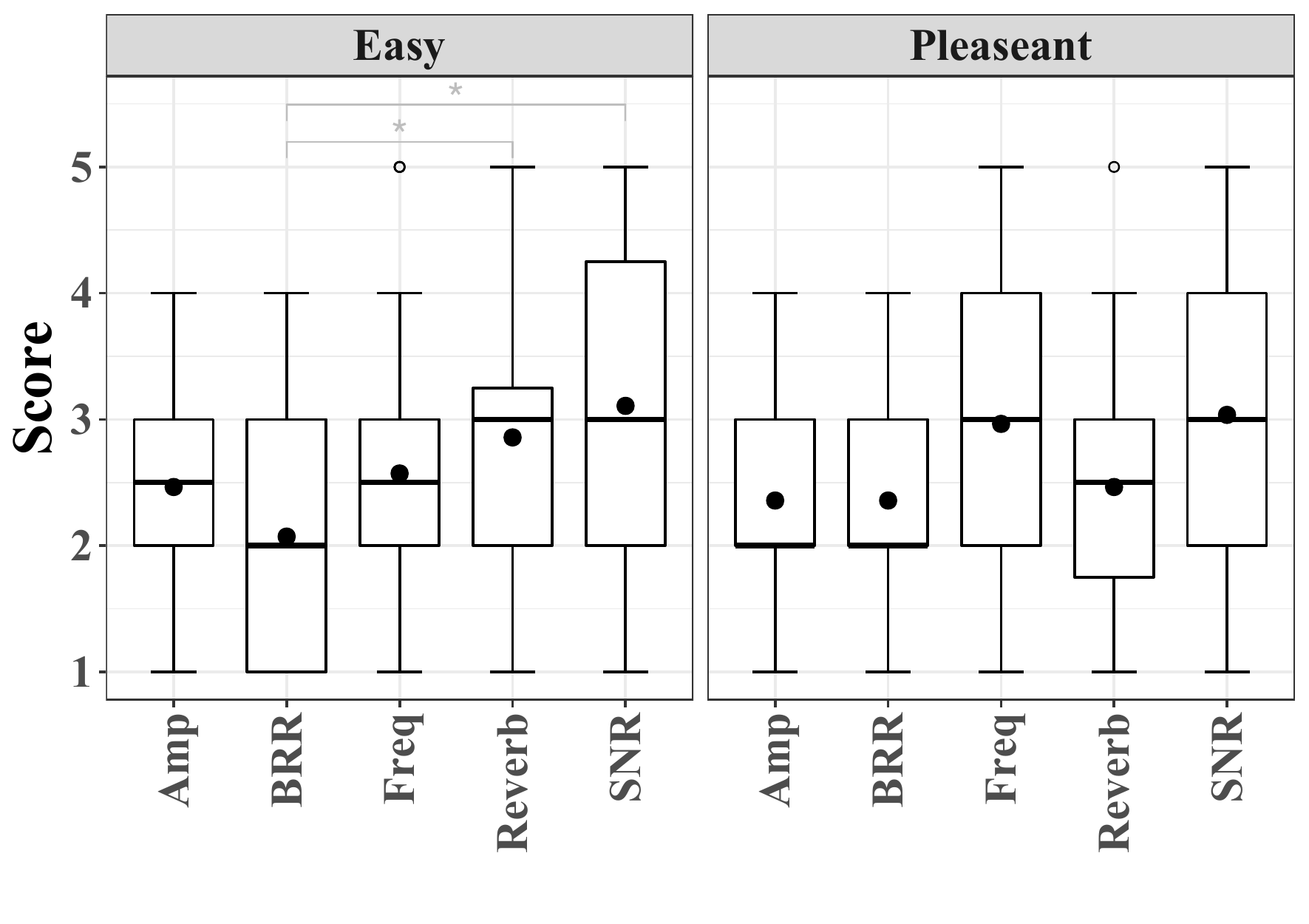} 
  \caption{Participants' responses to: left, ``Were sonifications easy to use to estimate depth positions?'' (score from 1: ``very easy'' to 5: ``very difficult''). Right, ``Were sonifications pleasant to listen to?  (score from 1: ``very pleasant'' to 5: ``very unpleasant'')}
      \label{fig:questEasePlease}   
\end{figure}

Figure~\ref{fig:questEasePlease} shows participants' answers to questions (\textit{Easy}) and (\textit{Plaseant}). A one-factor ANOVA showed a significant effect of the sonifications on answers to questions (\textit{Easy}) ($F_{4,108}=3.42, p=\num{0.011}$) and (\textit{Pleasent}) ($F_{4,108}=2.65, p=\num{0.037}$). Pairwise comparisons showed that~\textit{BRR} (Beep Repetition Rate) obtained a significantly better score than~\textit{Reverb} and~\textit{SNR} for the question (\textit{Easy}). This is not surprising since~\textit{BRR} led to overall better positioning accuracy than~\textit{Reverb} and~\textit{SNR}. For the question (\textit{Easy}),~\textit{Amp} scores were similar to~\textit{Freq} scores, and not significantly higher than~\textit{BRR} scores. Yet, the accuracy with~\textit{Amp} was overall worse than with~\textit{BRR} and~\textit{Freq}. Also, only 11 out of 28 participants (39\%) were more accurate with the sonification that they rated as the easiest to estimate depth positions. This suggests that a majority were not able to accurately accurately self-assess their performance. 

The lower median score with~\textit{Amp} and~\textit{BRR} for question (\textit{Pleasant}) suggest that individuals found these two latter sonifications more pleasant to listen to. 
  
Table~\ref{tab:mostNat} present answers to question (\textit{Natural}). ~\textit{BRR}~had the best results with 11 participants who found it to be the most natural sonification. We hypothesized that the most natural sonification for individuals was also the one that they found the easiest to estimate depth. Sixteen out of twenty-eight participants (57\%) responded that the most natural sonification was te one to which they gave the best score on question (\textit{Easy}). For 12 of them (43\%), there was therefore a difference between the most natural sonification and the easiest sonification. 
While~\textit{Freq} gave the best positioning errors in the first experimental stage, only 2 participants answered that it was the most natural sonification to encode depth. Finally, only 4 out of 28 (14\%) had their best performance with what they found to be the most natural sonification to encode depth. This also supports the idea that individuals were not able to accurately self-assess their performance. 


  
\begin{table}[]
\captionof{table}{Participants' answers to ``What sonification did you find the most natural to encode depth?''} 
\begin{tabular}{@{}cccccc@{}}
\toprule
                                                                                 & BRR                                                    & Amp                                                 & SNR                                                & Freq                                               & Reverb                                                  \\ \midrule
\begin{tabular}[c]{@{}c@{}}Number of part. for\\ question (Natural)\end{tabular} & \begin{tabular}[c]{@{}c@{}}11\\  (39.3\%)\end{tabular} & \begin{tabular}[c]{@{}c@{}}7\\  (25\%)\end{tabular} & \begin{tabular}[c]{@{}c@{}}7 \\ (25\%)\end{tabular} & \begin{tabular}[c]{@{}c@{}}2 \\ (7.1\%)\end{tabular} & \begin{tabular}[c]{@{}c@{}}1 \\ (3.6\%)\end{tabular} \\ \bottomrule
\end{tabular}
    \label{tab:mostNat}
\end{table}

Answers to question~\textit{Strategy} 
allowed us to analyze the learning strategies. 
 One was to memorize two or three audio encoded reference positions (e.g. closest, farthest, and middle position). 
Participants then used the memorized reference positions to find the target depth position by interpolation. 
The other was to move the box slowly and repeatedly across the depth range to hear the entire range of sound. 

 
 There were also strategies that depended on sonifications. 
With~\textit{BRR}, some participants tried to count the number of repetitions per second. 
With~\textit{Freq}, some used their knowledge of music to associate depths with musical notes. 

\section{Discussion}

%


 Many recent sensory substitution systems use audio-encoded depth. However, no study has yet focused on the encoding of depth with sound in such systems. 
Our protocol was designed to evaluate 5 different sonifications of depth
 without the need for visual feedback. 
Importantly, the sonification system can be set up with any camera and a standard laptop and could be easily adapted to investigate other depth sonifications. Twenty-eight sighted blindfolded individuals 
 completed positioning tasks using the 5 depth sonifications describe in section~\ref{sec:sonifDesign}.

All sonifications resulted in average errors that are under chance level. They are therefore all potentially viable for encoding depth in sensory substitution systems. 
Nevertheless, we observed significant differences that we discuss below. 

\subsection{Comparison with studies conducted with sighted non-blindfolded participants}
We compared our results to studies that have conducted research on the effectiveness of distance sonification, but with non-blindfolded participants. 
Those studies take place within the generic context of auditory display while we focus on sonification for sensory substitution. Despite the different contexts, we are interested in comparing the effectiveness of sonifications with and without visual feedback. This could also potentially help to better understand multi-sensory integration. 

In~\cite{Parseihian2016}, participants had to complete 
a one dimensional guidance task to a target location. 
The distance between the experimental object and a target is sonified. 
We sonify the distance between the user and the object. 
In one of their experimental setups, the authors compare the sonifications that we name~\textit{Freq},~\textit{BRR} (Beep Repetition Rate) and~\textit{Amp}. Despite the differences to our experimental protocol, they also found that~\textit{Freq} and~\textit{BRR} gave better accuracy than~\textit{Amp}.  

 

In~\cite{Parseihian2012}, the distance between the hand and a central reference point is sonified. Similar to our protocol, participants were asked to perform a positioning task after a learning phase. The authors compare the sonifications that we name~\textit{BRR},~\textit{Freq} and~\textit{Reverb}. Similar to our findings,~\textit{Reverb} gives worse accuracy than~\textit{Freq} and~\textit{BRR}. However, they found that~\textit{BRR} gives better performance than~\textit{Freq} while we found similar performance between these two sonifications. This could be explained by the use of a linear scale for~\textit{Freq} in their sonification design while we use a logarithmic scale (based on the Western music scale) which better fits to human perception of frequency. Moreover, they observe a strong tendency to overestimate small depths and underestimate large depths. We observe a similar but less significant effect. Indeed, 
the slopes of the linear regressions are closer to 1 (i.e. estimated depths were less biased) in our study. Zhang~\textit{et al.}~\cite{zhang2021} showed that estimated visual depths are subject to a ``contraction'' bias (i.e., overestimation of small depth and underestimation of large depth). Thus, it is possible that the bias exists also in the estimation and memorization of audio-encoded depths and is reinforced in~\cite{Parseihian2012} in which participants had visual feedback. 

In~\cite{Bazilinskyy2016}, the distance between a cursor and line on a computer screen is sonified. Participants were also asked to perform a positioning task after a learning phase. The authors compare the sonifications that we name~\textit{BRR},~\textit{Freq} and~\textit{Amp}. They found no difference in performance between the 3 sonifications while we found~\textit{BRR} and~\textit{Freq} to be significantly better than~\textit{Amp}. There can be several explanations. First, they used a linear scale for~\textit{Freq}, which could explain why~\textit{Freq} does not give a better accuracy than~\textit{Amp}. Moreover, the learning phase in their work was passive. It consisted in presenting 11 audiovisual stimuli corresponding to 11 positions equally distributed on the screen. There was no sensory-motor feedback loop, which is necessary for effective learning~\cite{Noe2006,maidenbaum2014sensory}. Thus, it seems that the differences in participants' accuracy in their work are mainly due to the memorization of audio-visual stimuli presented during learning. It could also explain why they did not observe differences between sonifications. 



\subsection{Additional findings}
\label{sec:addFind}

We found~\textit{BRR} to be the easiest sonification to remember. 
It is possible that with a longer learning period, the other sonifications would have been memorized in the same way. 
However, in accordance with Dakopoulos~\textit{et al.}~\cite{Dakopoulos2010}, 
a sensory subsititution device should not require a long learning time. \textit{BRR} therefore seems to be a promising sonification since it is well memorized without the need for a long learning period. Moreover, we found that~\textit{BRR} gives the best accuracy when depth and azimuth are simultaneously sonified.  

 With the exception of~\textit{Amp}, the sonifications resulted in estimates with low biases. The large estimation biases obtained with~\textit{Amp} can be explained by the work of Poulton~\cite{poulton1979models}, who shows that the quantitative judgment of loudness follows a contraction bias (i.e. an overestimation of soft sounds and underestimation of loud sounds). 
 
Qualitative feedback shows that a majority of participants found~\textit{BRR} to be the most natural and the easiest sonification to encode depth.~\textit{BRR} is present in everyday life and individuals are already used to this type of sound. 
A significant portion of the participants (25\%) found~\textit{SNR} to be the most natural way to encode depth. Participants may have naturally associated this sonification with the visual atmospheric perspective. 
Atmospheric perspective refers to the less detailed appearance of distant objects due to particles and air between the distant objects and the eyes. 
Also, a congenitally blind person with whom we had informal discussions about our depth sonifications noted that~\textit{SNR} gives two cues for estimating depth: the volume of noise and the volume of sinusoidal sound. It would be worthwhile to further investigate this sonification.

\subsection{Differences in effectiveness of sonifications}
\label{sec:SonifDiff}
The positioning accuracy is bounded by the smallest change in depth required to perceive differences in the sonification sounds. This is related to Just Noticeable Differences (JNDs) of each sonification, which are the smallest changes  in sound parameters that can be perceived by the auditory system. 
Parseihian~\textit{et al.}~\cite{Parseihian2016} suggest that JNDs might predict the effectiveness of sonifications (small JNDs would give a better effectiveness). 
We conducted an online audio experiment~\footnote{the experiment is available at: https://openprocessing.org/sketch/1536387} to estimate the minimum depth differences required to perceive distinct sounds. For each sonification, we first played a sound corresponding to a depth $d$ and we then played a sound corresponding to a depth $d+ r \cdot \Delta d$, with $r \in \{-1,0,1\}$. Parameter ranges of the online experiment are identical to those of the main experiment. 
Pairs of sounds were successively presented and listeners had to label the second sound as being different or equivalent to the first sound. $\Delta d$ was chosen according to a staircase procedure~\cite{StairCase1962}: if the participant's answer was correct, $\Delta d$ was reduced and if not, $\Delta d$ was increased. The procedure was stopped after 20 trials or when participants alternately answered right then wrong 5 times in a row. JNDs were computed as being the mean depth $d$ from the 5 last trials. Ten new listeners were recruited for the online experiment. 
We estimated JNDs for two depths $d=5~cm$ and $d=95~cm$. Results are in Table~\ref{tab:JND}. 
  
  \begin{table}[!htbp]
     \centering
        \captionof{table}{Minimum equivalent depth differences in $cm$ required to perceive two different sounds. Results are shown for depths of $95~cm$ and $5~cm$. }
 \begin{tabular}{cccccc}
    \cmidrule[\heavyrulewidth]{2-6}
                 		   & \textit{Amp}  & \textit{BRR} & \textit{Freq} & \textit{Reverb} & \textit{SNR} \\   \cmidrule[\heavyrulewidth]{1-6}
   $5~cm$ & $3.1\pm1.7$ &    $3.1\pm2.5$  &$1.3\pm1.9$&    $3.3\pm1.1$  &   $2.1\pm1.0$    \\ \cmidrule{2-6}
    $95~cm$& $6.7\pm1.8$ &    $2.2\pm1.5$   &$1.6\pm1.7$ &   $8.8\pm0.1$  &    $4.3\pm2.7$     \\ \bottomrule
  \end{tabular}
    \label{tab:JND}
  \end{table}	

 ~\textit{Freq} gives the smallest JNDs. 
 Therefore, it could explain why~\textit{Freq} gives better accuracy than the other sonifications. However, this does not explain why~\textit{BRR} and~\textit{Freq} resulted in similar accuracy. Also, while~\textit{Amp, Reverb} and~\textit{SNR} produced better JNDs around low depths, positioning errors with theses sonifications are higher or similar around low depths than around large depths (see the distribution of errors represented with grey barplots on Fig.~\ref{fig:linReg} page~\pageref{fig:linReg}). 
 This leads us to conclude that JNDs alone cannot explain differences between sonifications. 
Accuracy difference could also come from the natural association between depth perception and sound parameters, named cross-modal correspondences (CMCs). 
CMCs have already proven to be the basis for effective human machine interfaces~\cite{hamilton2016cross,Tsiros2017}. 
Some of these correspondences may emerge from regular exposure in everyday life~\cite{parise2013}.~\textit{BRR} is a sonification often used in our daily experience in alarm systems to warn of approaching dangers (e.g. in rear view cameras in cars). 
This could also explain why a good positioning accuracy is obtained with~\textit{BRR}. 
Differences could also arise from the potential of the sonifications to develop effective strategies. 
For example, some participants reported trying to count the number of repetitions per second with~\textit{BRR}, which may be easier to remember and to quantify than  other sound parameters. 



\subsection{Guidelines for depth sonification}
~\textit{BRR} (Beep Repetition Rate) gives a good accuracy with an unbiased representation of the depth, was the easiest to remember and has good qualitative feedback.~\textit{Freq} also gives good accuracy but is not as easy to remember and participants do not find it a natural sound to encode depth. 
Although the amplitude of a sound and its reverberations are natural cues used by the brain to estimate distance from the sound source, participants were less accurate with the sonifications~\textit{Amp} and~\textit{Reverb}. Moreover,~\textit{Amp} gives a biased representation of the depth and~\textit{Reverb} received poor qualitative feedback. Finally, even if~\textit{SNR} yields less accuracy than~\textit{BRR} and~\textit{Freq}, participants rated it as a natural sonification to encode depth and it gives a low bias representation of the depth. We therefore believe that this sonification should be investigated further and could be a viable design choice. 

Our results lead us to recommend using~\textit{BRR} for depth sonification. However it may not be possible to use~\textit{BRR} (e.g. if it is already used to encode another visual dimension). In this case, we would recommend to use~\textit{Freq} if the task requires accuracy. If not,~\textit{SNR} could also be a good alternative which feels natural for participants. Finally, we do not recommend~\textit{Amp} or~\textit{Reverb}. 
 
 \subsection{Limitations of the work}
 \label{sec:limitation}
The experiments were conducted with blindfolded sighted participants. Late blind people maintain their spatial abilities and spatial cognitive map from their early visual experiences~\cite{dormal2012}. They also adopt similar strategies to those used by sighted people when performing spatial tasks~\cite{ungar2000,Pasqualotto2013}. Therefore, we are confident that results should hold with late blind people. We also conducted a pilot study with a congenitally blind participant who had overall similar results to the blindfolded sighted participants (e.g. the congenitally blind participant was also more accurate with~\textit{BRR} and~\textit{Freq}). In the future, it would be interesting to compare our results with other experiments conducted with late-blind individuals and  with more congenitally blind individuals.

The five depth sonifications were evaluated for hand-reachable depths. Our results can therefore be used as such for systems which assist blind users to speed up the process of finding and grasping objects in front of them. 

Differences in accuracy between sonifications are possibly partially explained by the JNDs (see section~\ref{sec:SonifDiff}) of the sound parameters. 
Using the same sonification parameters over a larger depth range could therefore result in larger errors with a similar order of accuracy between sonifications. 
Beyond the hand reachable area, the accuracy is however less critical. The information would be used to either get into the hand reachable area of the object, or to avoid it. This is supported by Pressl and Wiesel~\cite{pressl2006computer}, who suggest that a SSVA device dedicated to navigation should encode the 
position of an object with an accuracy of 1 meter (or less) in order to reach it with a cane or with the hands.

In future works, it would be necessary to verify if sonfications (and particularly~\textit{BRR} which was the best for short depth range sonification) allow for accuracy of 1 meter or less over larger depths. It would also be interesting to compare the potential of the different sonifications to encode the depth of multiple objects at the same time.

\section{Conclusion}
Vision to audition sensory substitution devices that rely on depth sensors are a promising new tool for assisting blind people to sense their environment. A key challenge in the design of such systems is to define meaningful and intuitive association rules between visual features and sounds. In this work, we quantify the effectiveness of five sonifications of depth based on either loudness, reverberation, frequency, repetition rate of beeps or the signal-to-noise ratio of a pure tone mixed with white noise.

Overall, the repetition rate of  beeps is the best both quantitatively (positioning accuracy and memorability) and qualitatively (participants find it the most natural to encode depth). The sonification based on sound frequency provides good accuracy but participants did not find a natural fit to encode depth. The mixture of a pure tone with white noise received good qualitative feedback and could also be considered.

\appendices

\section{Computation of the chance level}
\label{annex:guessChances}
We compute the expected absolute postionning error under the assumption that participants would have estimated the object's position by chance (i.e. without using the sonification). Under this assumption, the target position as generated by the system $P_{t}$ and the position estimated by the participant $P_{p}$  would be uniformly distributed within the position range $[a;b]$ 
 and would follow the probability density function: 
\begin{equation}f_X (x) = \begin{cases} 1/(b-a) & \textrm{if} \quad{} x \in [a,b]\\ 0 & \textrm{otherwise,}\end{cases}\end{equation}
The absolute positioning error is: 
\begin{align} 
Err (P_{t},P_{p}) &=  |P_{t} - P_{p} | \\
 &=\begin{cases} P_{t} - P_{p}& \textrm{if} \quad{} P_{t} \geq P_{p}\\ P_{p} - P_{t} & \textrm{if} \quad{} P_{p} \geq P_{t}\end{cases}
\label{eq:absErr}
\end{align}


Still with the same assumption, we can assume that $P_{t}$ and $P_{p}$ are independent. We therefore compute the joint probability density function as:
\begin{align} 
f_{P_{t} P_{p}} (P_{t}, P_{p})&=f_{P_{t}}(P_t)\, f_{P_{p}}(P_p)\\
&=1/(b-a)^2
 \end{align}

The expectation of the positioning error is : 
\begin{align} 
\mathbb{E}(Err) &= \int_{a}^b\int_{a}^b Err(P_{t},P_{p}) \, f_{P_{t} P_{p}} (P_{t}, P_{p}) \,d P_{t} \, d P_{p}\\
&= \frac{1}{(b-a)^2} \int_a^b\int_a^b |P_{t} - P_{p}|  \,d P_{t} \, d P_{p}\\
 &= \frac{1}{(b-a)^2} \left[  \int_a^b\int_a^{P_{t}} (P_{t} - P_{p}) \,d P_{p} \, d P_{t} +\right. \\& \left. \int_a^b\int_{P_{t}}^b (P_{p} - P_{t}) \,d P_{p} \, d P_{t} \right]\\
&=  \frac{1}{(b-a)^2}\cdot \left[ \frac{(b-a)^3}{3}\right]\ = \frac{b-a}{3}
 \end{align}

\section*{Acknowledgment}
 NSERC-CRSNG for funding this research. The members of the NECOTIS research group for proofreading the paper. The participants, I. Balafrej, E. Richan and
the members of the NECOTIS research group for testing
and providing feedbacks. Fran\c cois C\^ ot\' e for the fruitful discussions on the blind community and providing feedbacks on the sonifications.

\ifCLASSOPTIONcaptionsoff
  \newpage
\fi




\bibliography{crossModal}
%

%
%
%






\end{document}